\documentclass[a4paper,11pt]{article}
\pdfoutput=1 

\usepackage{jinstpub} 

\usepackage{lineno}
\usepackage{caption}
\usepackage{subcaption}

\title{DTS-100G  - A versatile heterogeneous MPSoC board for cryogenic sensor readout}

\author[a,1]{T.\,Muscheid,\note{Corresponding author.}} 
\author[b]{A.\,Boebel,}         
\author[a]{N.\,Karcher,}        
\author[b]{T.\,Vanat,}          
\author[a]{L.\,Ardila-Perez,}   
\author[b]{I.\,Cheviakov,}
\author[a]{M.\,Schleicher,}
\author[b]{M.\,Zimmer,}
\author[a]{M.\,Balzer,}
\author[a]{O.\,Sander}          

\affiliation[a]{Karlsruhe Institute of Technology}
\affiliation[b]{Deutsches Elektronen-Synchrotron}

\emailAdd{timo.muscheid@kit.edu}

\abstract{Heterogeneous devices such as the Multi-Processor System-on-Chip (MPSoC) from Xilinx are extremely valuable in custom instrumentation systems. This contribution presents the joint development of a heterogeneous MPSoC board called DTS-100G by DESY and KIT. The board is built around a Xilinx Zynq Ultrascale+ chip offering all available high-speed transceivers using QSFP28, 28\,Gbps FireFly, FMC, and FMC+ interfaces. The board is not designed for a particular application, but can be used as a generic DAQ platform for a variety of physics experiments. The DTS-100G board was successfully developed, built, and commissioned. ECHo-100k is the first experiment which will employ the board. This contribution shows the system architecture and explains how the DTS-100G board is a crucial component in the DAQ chain.}


\proceeding{Topical Workshop on Electronics for Particle Physics\\
  Sept 19 - 23, 2022\\
  Bergen, Norway}

\begin{document}
\maketitle
\flushbottom

\section{Introduction}
\label{sec:intro}

Modern experiments in the field of particle physics use a large number of high-sensitive sensors for measuring the energy of particles, which enables further exploration of the universe~\cite{Gastaldo2017, Hamilton2022, Faverzani2016}. Since the measurement of particles often takes place at a timescale of micro or even nanoseconds, a very fine granular time resolution is essential for data acquisition. Furthermore, the observation time is often in the range of several months or even years. Developing adequate readout electronics suitable for these applications is a complex task. The system must be capable of simultaneously processing data from all sensors at a high sample rate. FPGAs are often used on applications requiring real-time signal processing to effectively reduce the data rate to downstream components. MPSoCs, including both the FPGA and several CPUs, can be deployed to increase the versatility of the readout system. The Programmable Logic (PL) on the device implements all deterministic workloads dealing with the data flow whereas the Processing System (PS) controls the communication with the board components and storage systems. 

The current generation of Xilinx MPSoCs features high-speed links with speeds of up to 32.75\,Gbps~\cite{XilinxZUSP} connected directly to the PL. Xilinx offers several evaluation boards built around these MPSoCs, but they only offer a limited subset of the MPSoC features. While these boards are suitable for prototyping, full-scale readout-electronics mostly requires custom hardware. 

If real-time data reduction is not possible on the frontend, the data must be transferred to a backend server. In recent years, several new interfaces for optical data transmission, such as FireFly~\cite{SamtecFirefly} and QSFP28, have been developed, which promise reliable, high-throughput board-to-board data transmission. These new optical transceivers enable transmission rates of up to 100\,Gbps and allow transfer of the data for post-processing and storage.
Within the Detector Technologies and Systems (DTS) program of Helmholtz~\footnote{https://www.helmholtz-detectors.de/}, DESY (Deutsches Elektronen-Synchrotron) and KIT (Karlsruhe Institute of Technology) developed the DTS-100G board as an universal and flexible platform. It functions as both a versatile MPSoC board for reading out various cryogenic sensors and an evaluation platform for high-speed optical transmission capabilities. This work aims to describe the board capabilities and its first application within the readout system for the Electron Capture of Ho\textsuperscript{163} experiment (ECHo), which investigates the upper limit of the electron neutrino mass.

\section{DTS-100G MPSoC board} 

\begin{figure}[h]
\centering
\includegraphics[width=.8\linewidth]{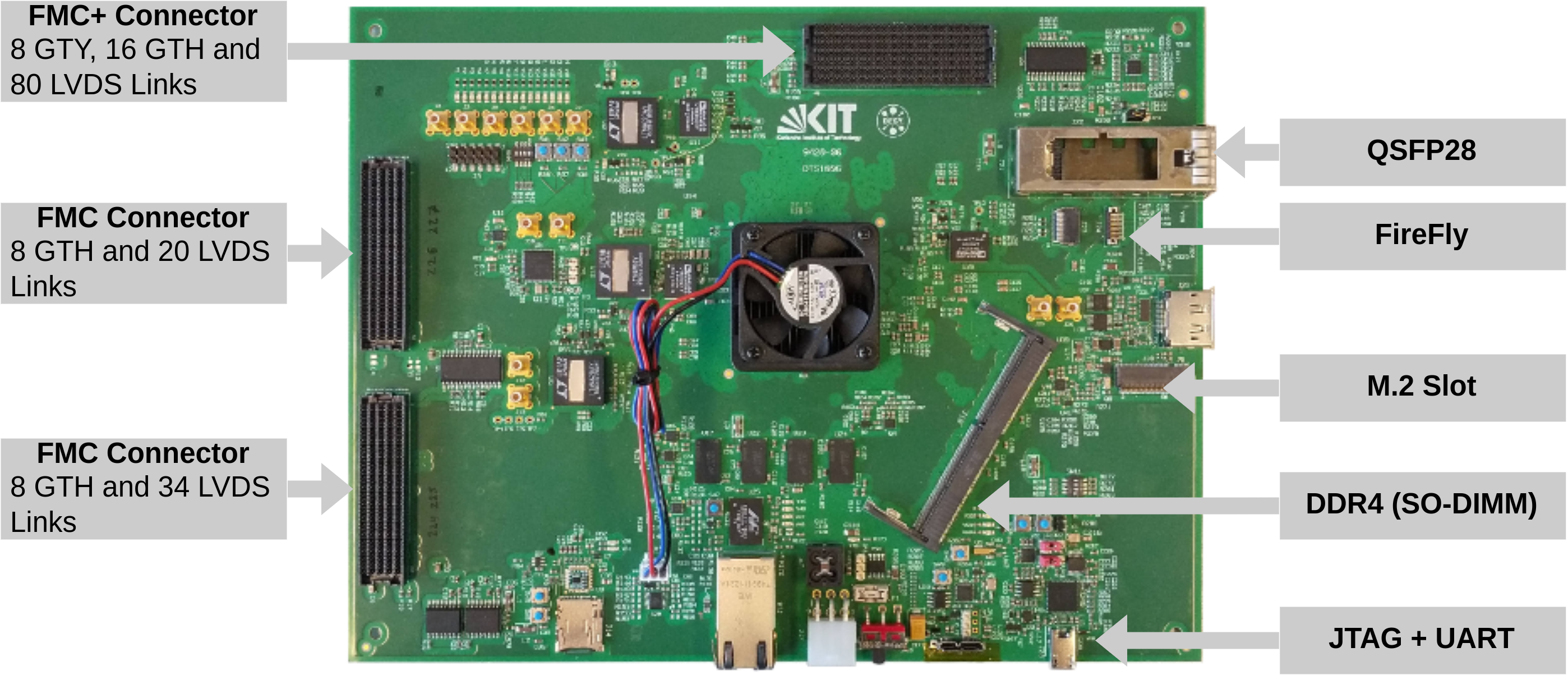}
\caption{DTS-100G MPSoC board.}
\label{fig:dts100g}
\end{figure}

The DTS-100G board, shown on Figure \ref{fig:dts100g}, has a dimensions of 258.5\,x\,198\,mm with a thickness of 2.56\,mm. It contains 18 signal and power layers. The board is built around a MPSoC from the Xilinx Zynq Ultrascale+ family. The first version of the board is equipped with a XCZU11EG-FFVC1760-2-E, but it can also be used with a footprint-compatible ZU17EG or ZU19EG device. The resources available in the PL of these MPSoCs are different, but the processing system is the same, so only small changes to the build system are needed to make a change. A second revision of the board is currently undergoing a few minor modifications and fixes.

The outstanding feature of the board is its connectivity. One of the development goals was to make use of all the MPSoC's high-speed links and provide them to users in a variety of connector types. The high-speed data transmission interfaces are directly accessible by the PL. The chosen MPSoC package offers 16 GTY and 32 GTH transceiver lanes with speeds of 32.75\,Gbps and 16.3\,Gbps respectively~\cite{XilinxZUSP}. The DTS-100G links all available high-speed lanes to various connectors. Four GTY transceivers of the MPSoC are connected to a FireFly connector and four to a QSFP28 connector, enabling 100\,Gbps board-to-board data transmission speed on each of these ports. The remaining eight GTY transceivers, as well as 16 GTH transceivers, are linked to a VITA 57.4 FMC+ connector for mezzanine card attachment. The other 16 GTH transceivers of the MPSoC can be accessed by two additional standard FMC connectors.

For data storage, the board offers a SO-DIMM slot with up to 16\,GB of DDR4 RAM connected to the PS and 8\,GB of on-board DDR4 RAM directly accessible from the PL. The processing system is connected to an M.2 connector that may be used for a SATA storage or as a PCIe link. Additionally, a Display Port, USB3.0, EEPROM, 2x UART over USB (FTDI), 1\,Gb Ethernet and 2x 1\,Gb QSPI flash memory are available. The clock for the board components is provided by a configurable jitter cleaner (SI5345). Configuration of the individual components is performed by I\textsuperscript{2}C and SPI buses from the MPSoC.

\section{Transceiver Characterization}

\begin{figure}[h]
\centering
\includegraphics[width=\linewidth]{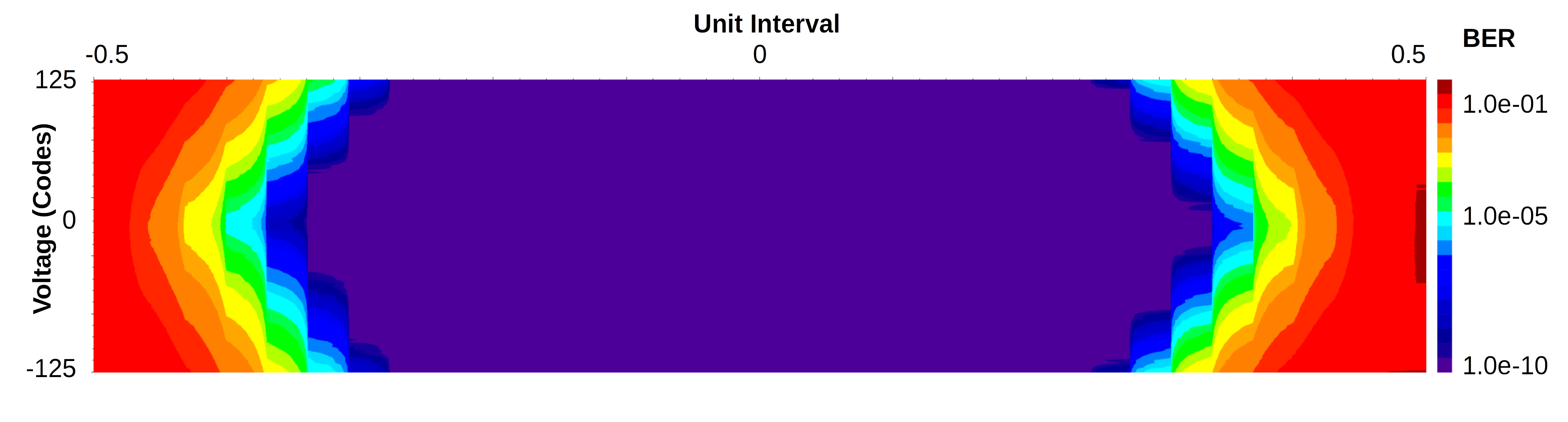}
\caption{Eye diagram of one FMC+ GTY-transceiver at 10 Gbps, other channels behave similarly. Measurement was performed using PRBS-31 with no pre- or post-emphasis compensation.}
\label{fig:eye}
\end{figure}

The board was designed for the use-case of acquiring measurement data from physical experiments, so the performance of the high-speed links received special attention during the design process. To verify the signal integrity of the high-speed links, Bit Error Rate (BER)~\cite{XilinxIBERT} tests were performed on the GTY-transceivers. Communication on the FMC+ connector was characterized by attaching a loopback FMC+ HSPC card by Samtec~\cite{samtec2019}, which generates a 156.25\,MHz reference clock and loops all high-speed links of the FMC+ connector as well as several low-speed links. For characterization of the signal integrity of FireFly and QSFP28, 0.0 dB loopback modules were attached to these connectors. Since the ports were operated in a low-attenuation loopback channel, the IBERT IP-Core was set to Low-Power Mode (LPM).
In the first step, the data rate for each transceiver was set to 10\,Gbps. Successful operation at this data rate is a key requirement for the first application of the DTS-100G within the ECHo experiment, which is introduced in section \ref{sec:echo}. Over several hours, $3 \cdot 10^{14}$ bits were generated and transferred by each transceiver. During this measurement, not a single bit error was detected, resulting in a BER less than $5 \cdot 10^{-15}$. Figure \ref{fig:eye} shows an example eye diagram for one of the transceivers with a BER of $1 \cdot 10^{-10}$. 

Subsequently, the transceivers were operated at 25\,Gbps, which is the maximum line rate the DTS-100G is designed for. Analyzing the IBERT results revealed that reliable communication in this edge case has not been completely achieved. On the FMC+ transceivers the test pattern had to be reduced from PRBS-31 to PRB-15. With this modification, no bit errors were detected during 24 hours of measurement with a transmission of $2.16 \cdot 10^{15}$ bits on each line. Communication via FireFly and QSFP28 was improved by additionally increasing the Pre- and Post-Cursor of the transmission signal. However, reliable communication on these links was only possible using PRBS-9 reaching a BER less than $1 \cdot 10^{-12}$ on both connections during the 24 hour measurement. Possible sources for the signal distortions at the 25\,Gbps line rate were identified on the PCB. The second revision of the board includes modifications to the ground layer below the transceiver pins and via structure of the transceivers to address these issues. However, the improvements still have to be tested after assembly of the revised board.


Nonetheless, because the transceivers already support 25\,Gbps line rates with a BER below $1 \cdot 10^{-12}$ for FireFly and QSFP28 lines, the board allows 100G Ethernet~\cite{IEEE802.3ba} communication. The four links of each connector operate at 25\,Gbps each, leading to a total data rate of 100\,Gbps. For testing the reliability of this protocol, the DTS-100G was connected to a Xilinx VCU108 Evaluation board~\cite{XilinxVCU108}. The VCU108 offers two separate QSFP28 ports, one of them was connected to the FireFly of the DTS-100G, while the other was attached to the QSFP28. Over the course of 64 hours, bidirectional UDP Ethernet packets with Reed-Solomon Forward Error Correction (RS-FEC) were sent. On each connection, more than $3 \cdot 10^{11}$\,packets, consisting of around 9000 bytes each, were transmitted, and not a single packet error was detected. An effective payload data rate of 99.20\,Gbps was achieved on both cases. 


\section{Integration into the ECHo experiment}
\label{sec:echo}

The ECHo experiment (Electron Capture of Holmium\textsuperscript{163}) aims to investigate the electron neutrino mass~\cite{Gastaldo2017}. It utilizes Metallic Magnetic Calorimeters (MMC)~\cite{Fleischmann2005}, a cryogenic sensor, to absorb and calculate the energy of X-ray photons emitted during the decay of Ho\textsuperscript{163} to Dy\textsuperscript{163}. By fitting the recorded energy spectrum of the photons against a modelled function, the maximum neutrino energy can be defined. In order to minimize thermal influence in the cryogenic environment, a microwave-SQUID-multiplexing~\cite{Kempf2014} principle is used, which allows readout of multiple sensors on a single line. One multiplexer chip contains 800 sensors with a total bandwidth of 4\,GHz between 4 and 8\,GHz. The concept for the readout electronics is described in~\cite{Sander2019}. Mixing of the readout tone between radio frequency and baseband as well as the digitization of the combs is performed by custom hardware~\cite{Gartmann2022}. The main part of the software-defined-radio system is an MPSoC board responsible for the generation of the tones as well as for processing of the modulated signals. The 4\,GHz comb is split into 5 subbands of 800\,MHz, where each complex subband consists of separate I and Q data streams. Conversion is performed at a clock rate of 1\,GHz by three four-channel DACs (AD9144) and five two-channel ADCs (AD9680) with internal digital down-conversion. In total, a conversion rate of 160\,Gbps in both directions is achieved. 
The third DAC is only used partially for generation of the readout tones, so the two unused channels could be used for the generation of a fluxramp signal, which is required for linearization of the SQUIDs~\cite{Mates2012}. However, this task is currently being carried out by an extension board with a slower DAC (MAX5898), which was developed for the first prototype of this system~\cite{KarcherDiss}. This LVDS-DAC operates at a sample rate of 125\,MHz and is able to convert two channels with 16\,bit each. 

Communication between the MPSoC board and the converters is realized by means of the JESD204B protocol, which encodes the data in the 8b/10b scheme and requires one GTY or GTH high-speed transceiver connection per effective byte. For the first prototype of the ECHo readout electronics containing only one of the five sub-bands, the Xilinx ZCU102 evaluation board was used. However, the full-scale system now requires 20 high-speed lanes with a bidirectional data rate of 10\,Gbps each, so this evaluation board is not suitable. Instead, the DTS-100G, with its FMC+ connector providing 24 high-speed links fits perfectly into this application. Since both the ZCU102 and the DTS-100G are built around an MPSoC chip of the Zynq Ultrascale+ family, adapting the firmware to the new board was simple. By adding the DTS-100G board support package including the device-tree and the configuration of the MPSoC into our custom build system based on Yocto~\cite{KarcherDiss}, the firmware could quickly be deployed and installed on the board.


\begin{figure}
     \centering
     \begin{subfigure}[b]{0.55\textwidth}
         \centering
         \includegraphics[width=\textwidth]{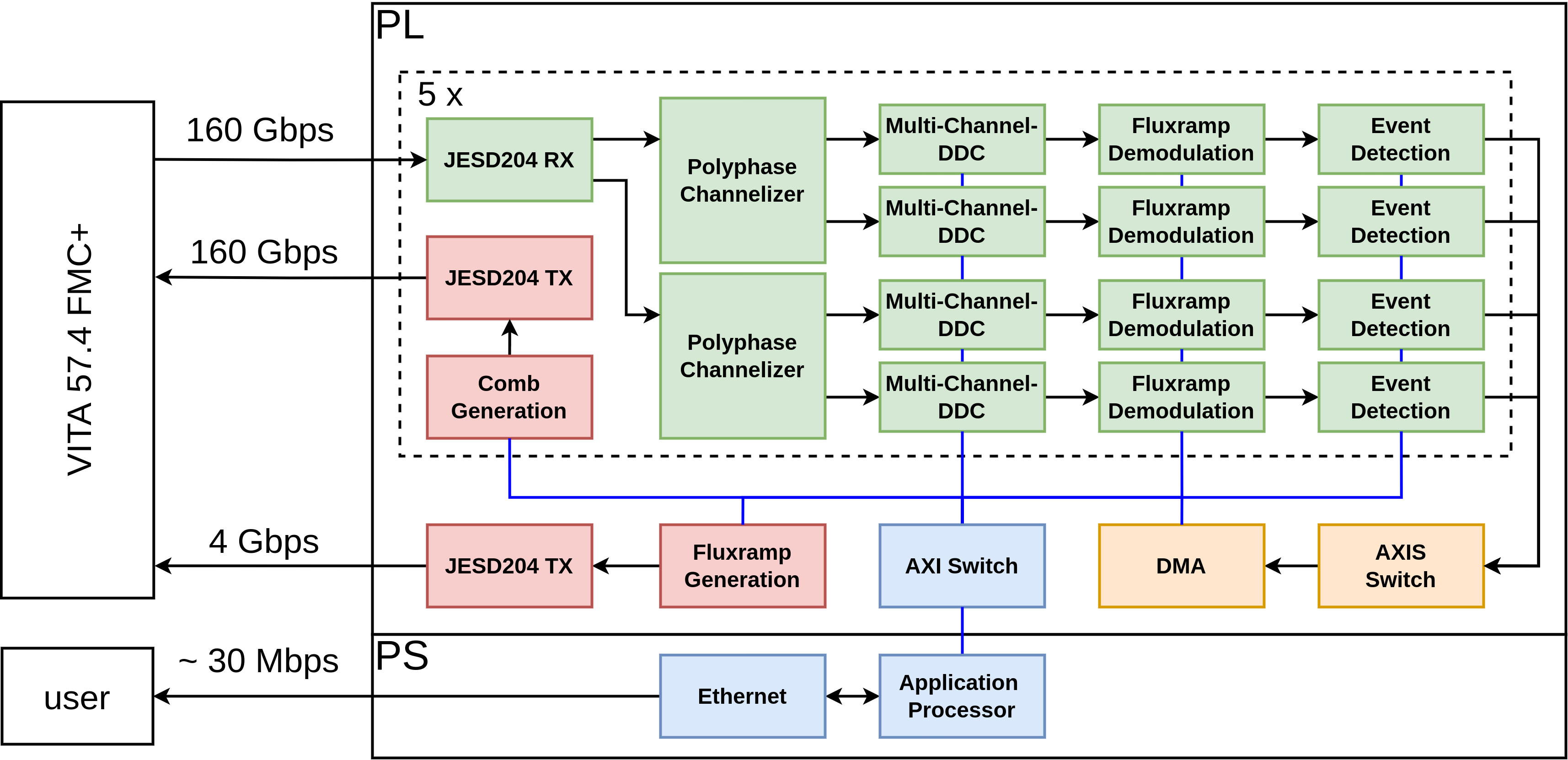}
         \caption{Full-scale firmware for readout of the ECHo experiment.}
         \label{fig:firmware}
     \end{subfigure}
     \hfill
     \begin{subfigure}[b]{0.4\textwidth}
         \centering
         \includegraphics[width=\textwidth]{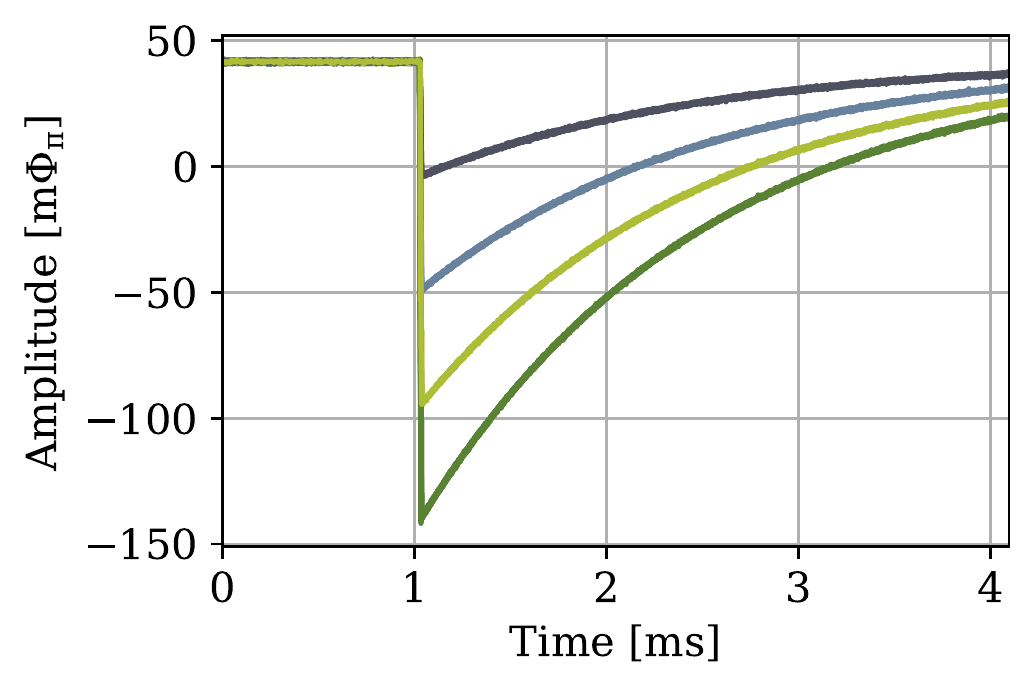}
         \caption{Superimposition of extracted pulses.}
         \label{fig:data_output}
     \end{subfigure}
     \hfill
    \caption{Usage of the DTS-100G as part of the ECHo electronics.}
    \label{fig:echo}
\end{figure}

The processing chain in the PL consists of several stages to minimize the data rate step by step. After separating the channels from each other and demodulating the data in order to reconstruct the raw sensor signal, the samples containing information about the decay energy are extracted, while the samples during idle state are discarded~\cite{Karcher2022}. Each 800\,MHz subband is handled by a separate chain. Processing of the full bandwidth therefore requires five parallel instances of the entire chain. The parallel chains are merged by a switch module, that combines the individual signals to one single data stream. A general overview on the firmware is given in Figure \ref{fig:firmware}. At the output, the data rate to be stored on the DDR4 or transmitted to a server is around 240\,Mbps, depending on the exact decay rate of the isotope. The extracted pulses are stored in packages with meta information and up to 1024 samples. Figure \ref{fig:data_output} shows a superimposition of extracted samples from multiple detected pulses with four different energy levels. Due to the data rate minimization, the standard 1\,Gb Ethernet connection is sufficient to send the data packages for post processing. However, high-speed data transfer might be used in the future in case access to raw data is desired, such as for system calibration. 

\section{Summary}

The DTS-100G MPSoC board is a universal platform suitable for the readout of particle physics detectors. The FMC+ connector, with 24 high-speed links and several 100\,Gbps data transmission interfaces, allows acquisition of measurement data both with a high sampling rate and a high bit-width for maximum sensitivity. Characterization of the GTY transceivers showed error-free data transmission at 25\,Gbps and effective data rates of over 99\,Gbps on both four-lane FireFly and QSFP28 interfaces.

The first application of the DTS-100G will be the ECHo-100K experiment, which investigates the upper limit of the neutrino mass. The DTS-100G, as part of the readout electronics, is responsible for the generation of 400 readout tones and real-time processing of the modulated signals. In total, 160\,Gbps of bidirectional data is handled by the transceivers. Since the DTS-100G provides 24 high-speed links on an FMC+ connector, it is perfectly suitable for this application. The PL within the MPSoC offers sufficient flexibility for implementation of the required processing chain. A full vertical slice of the ECHo readout electronics is currently under commissioning and characterization in preparation for the production of 15 platforms required by the experiment.  





\bibliographystyle{unsrturl}
\bibliography{dts100g}

\begin{thebibliography}{10}

\bibitem{Gastaldo2017}
L.~Gastaldo et~al.
\newblock {The electron capture in 163Ho experiment -- ECHo}.
\newblock {\em {The European Physical Journal Special Topics}},
  226(8):1623--1694, Jun 2017.
\newblock \href {https://doi.org/10.1140/epjst/e2017-70071-y}
  {\path{doi:10.1140/epjst/e2017-70071-y}}.

\bibitem{Hamilton2022}
The~QUBIC collaboration et~al.
\newblock {QUBIC I: Overview and science program}.
\newblock {\em Journal of Cosmology and Astroparticle Physics}, 2022(04):034,
  Apr 2022.
\newblock \href {https://doi.org/10.1088/1475-7516/2022/04/034}
  {\path{doi:10.1088/1475-7516/2022/04/034}}.

\bibitem{Faverzani2016}
M.~Faverzani et~al.
\newblock {The HOLMES Experiment}.
\newblock {\em Journal of Low Temperature Physics}, 184(3):922--929, Aug 2016.
\newblock \href {https://doi.org/10.1007/s10909-016-1540-x}
  {\path{doi:10.1007/s10909-016-1540-x}}.

\bibitem{XilinxZUSP}
Xilinx.
\newblock {DS891, Zynq US+ MPSoC DS: Overview}.
\newblock v1.9, 2021.

\bibitem{SamtecFirefly}
Samtec.
\newblock {Firefly: ECUE Series in 28 Gbps (OIF CEI-28G-VSR) Applications}.
\newblock 2013.

\bibitem{XilinxIBERT}
Xilinx.
\newblock {PG196, IBERT for UltraScale GTY Transceivers}.
\newblock v1.2, 2021.

\bibitem{samtec2019}
Samtec.
\newblock {VITA 57.4 FMC+ HSPC/HSPCe LOOPBACK CARD}.
\newblock 2019.

\bibitem{IEEE802.3ba}
IEEE~Computer Society.
\newblock {802.3ba, Part 3: Carrier Sense Multiple Access with Collision
  Detection (CSMA/CD) Access Method and Physical Layer Specifications}.
\newblock 2010.

\bibitem{XilinxVCU108}
Xilinx.
\newblock {UG1066, VCU108 Evaluation Board User Guide}.
\newblock v1.5, 2019.

\bibitem{Fleischmann2005}
A.~Fleischmann, C.~Enss, and G.M. Seidel.
\newblock {\em {Metallic Magnetic Calorimeters}}.
\newblock Springer Berlin Heidelberg, 2005.
\newblock \href {https://doi.org/10.1007/10933596_4}
  {\path{doi:10.1007/10933596_4}}.

\bibitem{Kempf2014}
S.~Kempf, M.~Wegner, L.~Gastaldo, A.~Fleischmann, and C.~Enss.
\newblock {Multiplexed Readout of MMC Detector Arrays Using Non-hysteretic
  rf-SQUIDs}.
\newblock {\em Journal of Low Temperature Physics}, 176(3):426--432, Aug 2014.
\newblock \href {https://doi.org/10.1007/s10909-013-1041-0}
  {\path{doi:10.1007/s10909-013-1041-0}}.

\bibitem{Sander2019}
O.~Sander, N.~Karcher, O.~Krömer, S.~Kempf, M.~Wegner, C.~Enss, and M.~Weber.
\newblock {Software-Defined Radio Readout System for the ECHo Experiment}.
\newblock {\em IEEE Transactions on Nuclear Science}, 66(7):1204--1209, 2019.
\newblock \href {https://doi.org/10.1109/TNS.2019.2914665}
  {\path{doi:10.1109/TNS.2019.2914665}}.

\bibitem{Gartmann2022}
R.~Gartmann, N.~Karcher, R.~Gebauer, O.~Kr{\"o}mer, and O.~Sander.
\newblock {Progress of the ECHo SDR Readout Hardware for Multiplexed MMCs}.
\newblock {\em {Journal of Low Temperature Physics}}, Sep 2022.
\newblock \href {https://doi.org/10.1007/s10909-022-02854-1}
  {\path{doi:10.1007/s10909-022-02854-1}}.

\bibitem{Mates2012}
J.~A.~B. Mates, K.~D. Irwin, L.~R. Vale, G.~C. Hilton, J.~Gao, and K.~W.
  Lehnert.
\newblock {Flux-Ramp Modulation for SQUID Multiplexing}.
\newblock {\em Journal of Low Temperature Physics}, 167(5):707--712, Jun 2012.
\newblock \href {https://doi.org/10.1007/s10909-012-0518-6}
  {\path{doi:10.1007/s10909-012-0518-6}}.

\bibitem{KarcherDiss}
N.~Karcher.
\newblock {\em {Ausleseelektronik für magnetische Mikrokalorimeter im
  Frequenzmultiplexverfahren}}.
\newblock PhD thesis, Karlsruher Institut für Technologie (KIT), 2022.
\newblock \href {https://doi.org/10.5445/IR/1000148040}
  {\path{doi:10.5445/IR/1000148040}}.

\bibitem{Karcher2022}
N.~Karcher, T.~Muscheid, T.~Wolber, D.~Richter, C.~Enss, S.~Kempf, and
  O.~Sander.
\newblock {Online Demodulation and Trigger for Flux-ramp Modulated SQUID
  Signals}.
\newblock {\em Journal of Low Temperature Physics}, Sep 2022.
\newblock \href {https://doi.org/10.1007/s10909-022-02858-x}
  {\path{doi:10.1007/s10909-022-02858-x}}.

\end{thebibliography}

\end{document}